\documentclass{article}

\usepackage{mathptmx}       
\usepackage{helvet}         
\usepackage{courier}        
\usepackage{type1cm}        
\usepackage{makeidx}         
\usepackage{graphicx}        
\usepackage{multicol}        
\usepackage[bottom]{footmisc}

\usepackage{bm}
\usepackage{amsmath,amssymb}
\newcommand\independent{\protect\mathpalette{\protect\independenT}{\perp}}
\def\independenT#1#2{\mathrel{\rlap{$#1#2$}\mkern2mu{#1#2}}}
\makeindex             

\usepackage{hyperref}
\newcommand{\footremember}[2]{%
  \footnote{#2}
    \newcounter{#1}
    \setcounter{#1}{\value{footnote}}%
}
\newcommand{\footrecall}[1]{%
    \footnotemark[\value{#1}]%
} 

\title{A Bayesian Analysis of Migration Pathways using Chain Event Graphs of Agent Based Models}

\author{Peter Strong \footremember{complexity}{Centre for Complexity Science, University of Warwick, Coventry CV4 7AL, UK.} \footremember{turing}{The Alan Turing Institute, British Library, 96 Euston Road, London NW1 2DB, UK.}  \\ \href{mailto:P.R.Strong@warwick.ac.uk}{P.R.Strong@warwick.ac.uk} \and Alys McAlpine \footremember{lshtm}{Gender Violence and Health Centre, Faculty of Public Health and Policy, London School of Hygiene and Tropical Medicine, London WC1E 7HT, UK.}\\ \href{mailto:Alys.McAlpine@lshtm.ac.uk}{Alys.McAlpine@lshtm.ac.uk}
\and Jim Q Smith \footremember{stats}{Department of Statistics, University of Warwick, Coventry CV4 7AL, UK} \footrecall{turing}\\ \href{mailto:J.Q.Smith@warwick.ac.uk}{J.Q.Smith@warwick.ac.uk}}

\begin{document}


%
%
\maketitle



\abstract{Agent-Based Models (ABMs) are often used to model migration and are increasingly used to simulate individual migrant decision-making and unfolding events through a sequence of heuristic if-then rules. However, ABMs lack the methods to embed more principled strategies of performing inference to estimate and validate the models, both of which are of significant importance for real-world case studies. Chain Event Graphs (CEGs) can fill this need: they can be used to provide a Bayesian framework which represents an ABM accurately. Through the use of the CEG, we illustrate how to transform an elicited ABM into a Bayesian framework and outline the benefits of this approach.}

\section{Introduction}
\label{Sec:1}
Researchers and policymakers are interested in modelling migration as they aim to understand the mechanisms involved in order to inform policy. For example, organisations may aim to promote safe labour migration in line with the UN's Sustainable Development Goals \cite{united_nations}. Migration can increase vulnerability to human trafficking and exploitation. It is estimated that 23\% of victims of forced labour \cite{report:} and  60\%  of victims of human trafficking were outside their country of residence \cite{mon_2017}. In order to inform policymakers attempting to prevent exploitation, it is important to understand migrants' journeys and identify how individuals' hyper-precarity and livelihood insecurity, experienced due to both employment and immigration \cite{doi:10.1177/0309132514548303}, evolves on different migration pathways.

Increasingly, Agent Based Models (ABMs) have been commonly used in contexts such as migration as they focus on the level of the individual and can be constructed by modelling the potential outcomes of successive events and decision-making \cite{mcalpine_kiss_zimmerman_chalabi_2020,klabunde}. In order to construct these models, a range of data sources, such as large structured demographic datasets or natural language narratives and theories have been used to inform deterministic and stochastic transitions within an ABM. These transitions take the form of either mathematical equations, such as differential equations, or heuristic if-then rules and are informed by experts who describe the influences, possible options available and threats along their journey to another country. However there is often ambiguity in the reporting of these models.  For this type of egocentric modelling with heterogeneous actors and actions, where the focus is on an individual's decisions, ABMs are an obvious choice and hence are being increasingly applied to model migration, though not yet with great detail on the true range of actors and decisions due to the complex nature of the application and difficulty in acquiring testimonies. Despite their increasing popularity, ABMs are unable to naturally combine expert judgement with available data to estimate and validate them. This is a problem as these steps are particularly important in this domain due to the previously mentioned difficulty in obtaining large amounts of data.

Chain Event Graphs (CEGs) are directed acyclic graphs that describe the evolution of a process through an unfolding of events \cite{SMITH200842}. CEGs are transformations on event trees and therefore are able to represent context-specific independence statements, conditional independence statements that are true only in specific contexts. The CEG should be thought of as a collection of \textit{florets} (non-leaf nodes and their outgoing edges) that represent the events and their outcomes of the modelled process. The CEG represents the aforementioned independence statements by providing a staging on the florets that denotes their exchangeability. An Example demonstrating these concepts is shown in Section \ref{Sec:4}. A particular class of CEGs, non-stratified CEGs, are able to more naturally represent an asymmetric unfolding of events. More generally, CEGs have previously been used for modelling in a wide range of applications, such as criminal collaborating \cite{bunnin2020network}, public health \cite{shenvi2019bayesian} and educational studies \cite{freeman_smith_2011}.

In this paper, we present a new methodology being developed to provide a Bayesian framework to an existing ABM through transforming it into a CEG. There are many key benefits of transforming the ABM into a CEG. One key advantage of the CEG is its compact representation, which not only shows the asymmetries in the events, as was the case with the initial diagram of the ABM, but also explicitly represents the context-specific conditional independences within the graph's topology. As a result, the potential series of events that may be experienced by a migrant and how these events impact future events are easily comprehensible. Secondly, by using the transformation of an ABM into a CEG we can apply a Bayesian framework in a natural way. This is particularly valuable in the situation where, due to the nature of migration data, the ability to perform Bayesian inference to combine data expressed through individual testimonies or expert descriptions is vital. Further benefits include the ability to use Bayesian model selection to compare the likelihood of different independence statements around the outcomes of events, represented by different theories of migration, using Bayes factor. For these reasons, the CEG makes a highly effective conduit into a stochastic description of the problem. 



This is the first paper that investigates how an ABM can be used to construct a CEG; it is the first genuine Bayesian model of migration processes to be built which draws from a combination of testimonies, surveys typical data and expert judgement. In Section \ref{Sec:2}, we give a background into ABMs of migration and formalise the class of models we are considering. In Section \ref{Sec:3}, we introduce the CEG, explain how it can represent the ABM and the benefits of this approach. In Section \ref{Sec:4}, we provide an example of how to convert a given ABM into a CEG. We conclude with a discussion of future work.

\section{Agent Based Models of Migration}
\label{Sec:2}

Migrants' pathways are often complex and non-linear, making many conventional modelling approaches unsuitable. ABMs provide a bottom-up approach to modelling, where the focus is on the individual. The aim of these models is to accurately replicate a population, its environment and the interactions that occur.

Despite their ability to plausibly model the transitions of an agent, many ABMs, both in migration research \cite{mcalpine_kiss_zimmerman_chalabi_2020} and more broadly \cite{math_abm_hinkelmann_murrugarra_jarrah_laubenbacher_2010, grimm_berger}, have been described as opaque with many of the critical details needed to fully understand or replicate the models missing from publication, such as the lack of standardising model development. Some attempts, such as the ODD protocol, have been made to create a standardised structure for explaining ABMs \cite{grimm_berger}, but there is still significant variance in how the protocol is used and the clarity it brings to ABMs. ABMs' application often depends on the implementation of often severely constraining software which may or may not match the modelled domain well. Perhaps even more concerning is the gulf that exists when applying such models between the domain and a principled statistical inferences about that domain. In particular no real guidance about how to set the ABM parameters is given, estimation of these is naive and model selection performed simply by matching trajectories of hypothesised models with chosen/estimated parameters with sampled trajectories. As a result, others \cite{grimm_abmbad,heckbert_abmbad,schulze_abmbad,an_abmbad} have already identified the desperate need for embedding more principled ways of performing inference in order to estimate and validate ABM models when these are applied to real case studies. In this paper we argue that the best way of doing this is by using Bayesian models formulated around tree based CEG methods in ways we illustrate below. 

As a first step we of course need to provide a proper formal systematic description of an ABM -- something that is sadly missing from many applications of this promising technology. Here we follow \cite{math_abm_hinkelmann_murrugarra_jarrah_laubenbacher_2010} who express the ABM as a particular class of dynamic system model where agents are variables and their transitions are given by local updating functions. This work provides a similar statistical framework in order to study ABMs. We consider a set of agents $(x_1,x_2,\dots,x_n)$ that take values in $\mathbb{S}$ a finite discrete set that represents the possible states that an agent can be in. The set of all possible values of all of the agents in the system gives the state-space. For any given state in the space, the updating process that determines the transitions between states is a Markov process. The possible transitions in the Markov process can be represented by a directed graph $G=(V,E)$ with $V$ the state space and edges $e\in E$ between $u\in V$ and $v\in V$ if it is possible to transition from state $u$ to $v$.

To provide a comprehensive translation of general ABMs as formally described above into Bayesian stochastic models would be a massive task and beyond the scope of this short paper. Here, for simplicity, we constrain our discussion to those ABMs with only one agent, and with a Markov process that has graph representation in the form of a finite, rooted, directed tree. The simplification of only using one agent is reasoned by the nature of these models being largely egocentric with the process and decision-making depending solely on the state of the individual, even if affected by interactions with other agents and the environment. The rationale of only allowing a finite, rooted, directed tree for the updating of states is justified: due to the nature of models of migrants pathways, we are interested in ABMs that can be thought of as an unfolding of a sequence of events. A tree gives the most natural representation of this process \cite{shafer_1996}.

\section{From ABMs to CEGs}
\label{Sec:3}

In Section 2, we described the class of ABMs that we are considering in this work. This decision was justified as the types of information we have about this process is best represented through a probability tree representing the possible progress of each migrant in the migrant population. This is particularly useful as it depicts the step-by-step nature of the process, where each migrant decides their next course of action. Typical hypotheses concerning this progress assume various conditional independence hypotheses, such as those shown in Section 4. Within an event tree model, these can be expressed by the stage structure on the florets of the tree. 

The if-then rules within a heuristic, egocentric ABM implicitly include independence hypotheses regarding the outcome of an event for an individual through the choice of inputs considered. By assuming this conditional independence within a hypothesised model, we can identify those migrants within a sample who can be assumed on the next step of their journey to be exchangeable with each other. The CEG provides a framework in which to embed this model. 

Bayesian methods are critical within such models because whenever models are sufficiently large to give a credible description of the processes, many parts of such processes are only sparsely observed. It is, therefore, critical to embed expert judgements through the priors on the hyperparameters. (In this work, this is the distributions on the prior floret probabilities). In this way, our proposed methodology scales up to granularities of descriptions shared by the ABMs of such processes.

We can embed not only the  prior expectations of these probabilities - as often needed in typical ABMs - but also their uncertainty. This embellishment means that, by using the exchangeability assumptions alluded to above and embedded in a Bayesian model, we can perform a prior-to-posterior update on these probabilities. In particular, we can derive principled model selection algorithms that respect the relative security of knowledge of different transitions within the system. We note that, even if no actual steps in some of the paths are observed, we can proceed with this inference, whilst if many people are observed making a particular collection of transitions then estimated transition probabilities will be close to their sample proportions. The model is suitably regularised.

Furthermore, if we assume floret independence, we can perform a conjugate Bayesian analysis (for full details, see \cite{freeman_smith_2011} and \cite{collazo}). The consequent Bayesian model estimation and selection is both transparent and rapid due to the closed form representation and the interpretative understanding of the hyperparameters. 

In particular, assuming each transition is multinomially distributed over the set of outcomes, to perform a conjugate analysis, we need to set the Dirichlet priors. The distributions for the transition probabilities are often not elicited in advance, due to the non-Bayesian nature of ABMs. However, if the values elicited are the mean transition probabilities, we can use these values as the prior means for the Dirichlet prior. In order to get the full prior distribution, we must add in a count of effective sample size. This acts as a measure of strength of the beliefs held within the ABM. This can be done either by eliciting such a value or by completing a sensitivity analysis around the value chosen, similar to the method taken in \cite{shenvi2019bayesian}. Other methods for setting up the hyperparameters can be seen in \cite{collazo}.

In order to compare competing models we can set the hyperparameters so they match each other as closely as possible as in \cite{heckerman}. This is done through a mind experiment where strengths of opinions are expressed concerning phantom samples over potential root to leaf path developments. 

Of course, we could fit a CEG directly to model the migration process, through eliciting an event tree, the hypothesises and the prior distributions. However, if such an ABM has already been developed and thoughtfully calibrated to domain understanding -- as is often the case -- then it would be inefficient to ignore this information. As we can exploit the fact that the CEG is largely compatible with the ABM, it can be used to embellish the original, rather coarse, description given by the ABM into an inferential model which is fit for purpose.

\section{An Illustrative Example of Migrant Behaviour}
\label{Sec:4}

\begin{figure}[h]
\centering
\includegraphics[width=\textwidth]{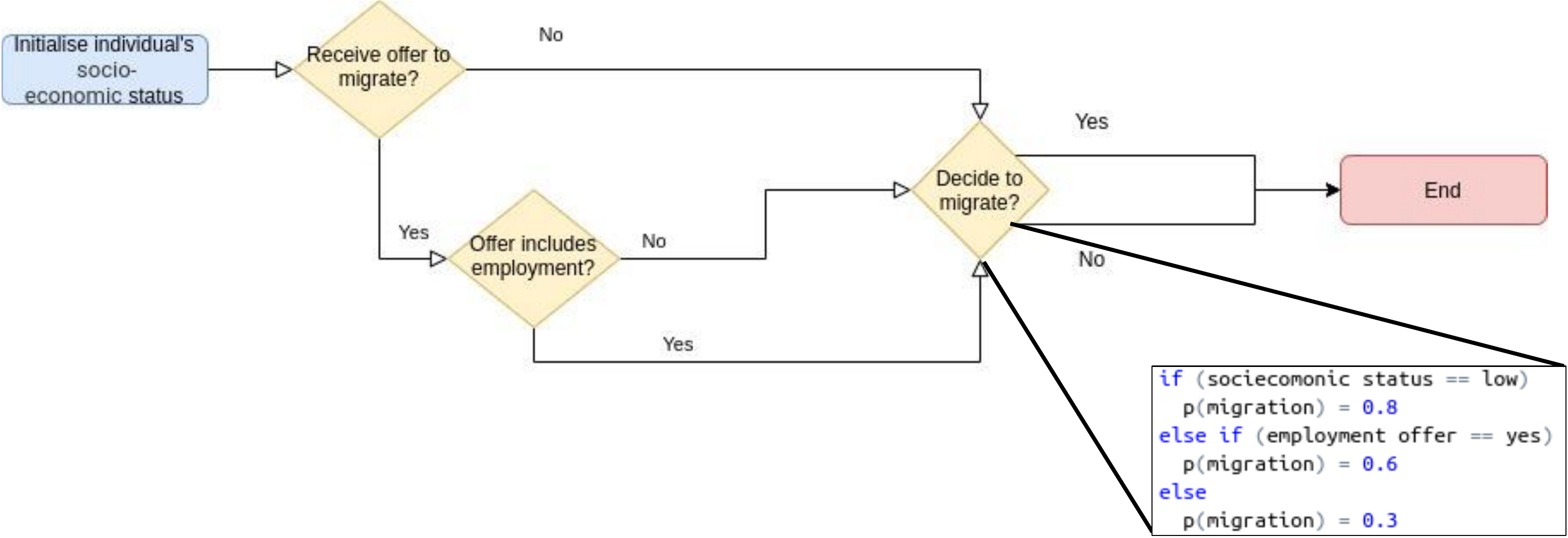}

    \caption{Example of an agent based model for migration.}
    \label{ABM_example}
\end{figure}

In this section, we will provide an illustrative example of the methodology described in the previous section. We will demonstrate this for the ABM represented in Figure \ref{ABM_example}. This ABM represents an individual's decision on whether to migrate through a sequence of events that impact their final decision. In this example, the ABM starts by initialising an individual's socio-economic status, $X_I$. The individual then may receive an offer to migrate, $X_O$. This offer either comes with or without employment, $X_E$. Finally, the individual makes a decision as to whether they should migrate or not, $X_M$. Each of the nodes in this diagram has an if-then rule associated with its transitions. For instance, Figure \ref{ABM_example} shows an example heuristic rule for the decision to migrate. This rule shows how the probability of migrating is dependent on the outcomes of previous events.

\begin{figure}[h]
\centering
    \includegraphics[width=\textwidth]{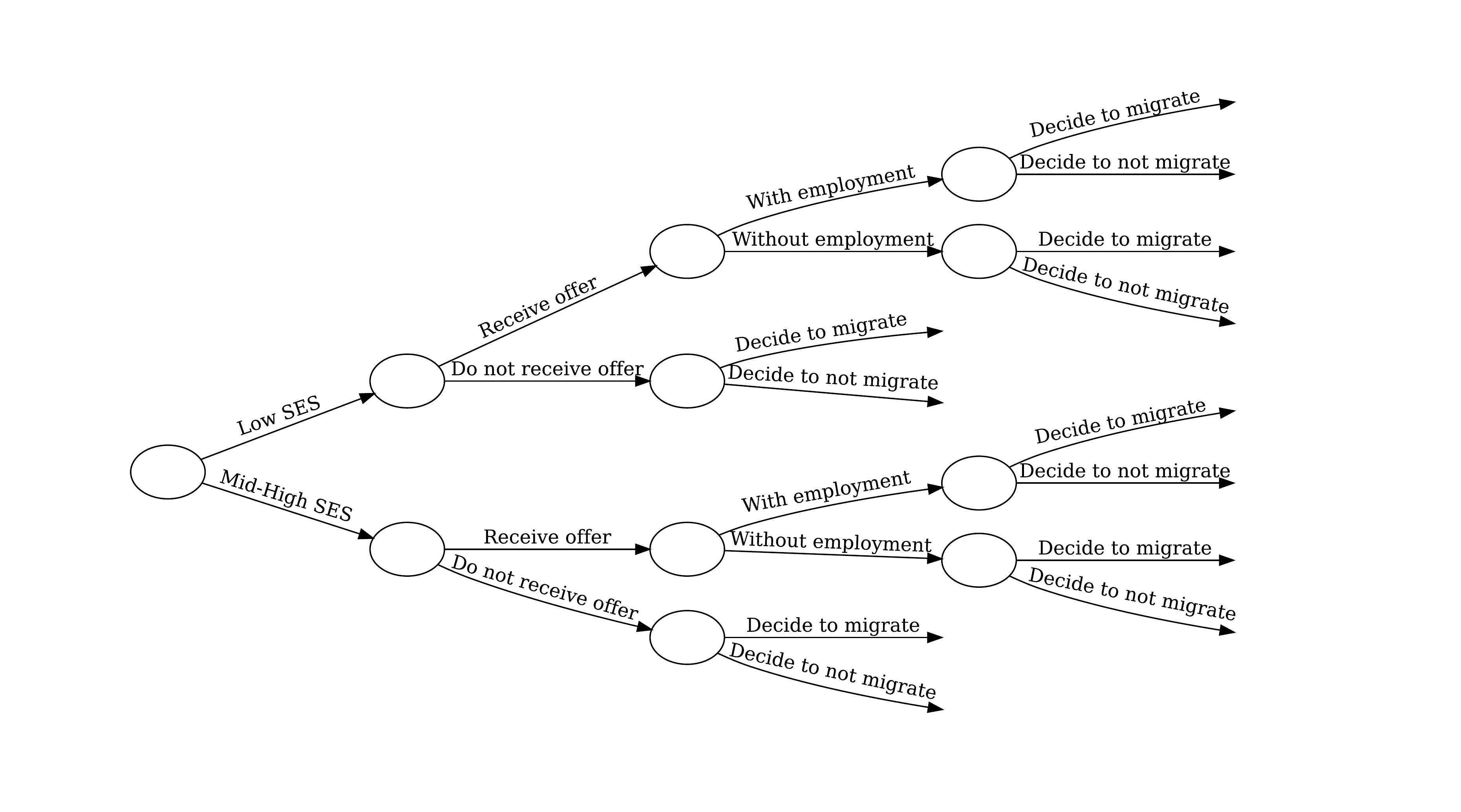}
    \caption{Event tree representation of the ABM shown in Figure 1. The leaf nodes are suppressed to prevent visual cluttering.}
    \label{tree}
\end{figure}

By untangling the current representation, we can obtain an event tree which is implied by the ABM. Within this class of ABM, an agent's transitions are determined by the outcomes of their previous transitions. Therefore, the next transition is conditional on its previous events. Such events define the situations (non-leaf nodes) in the CEG; there is a direct link between the CEG and the ABM. The nodes in the ABM define the situations in the CEG, with the possible transitions from that node represented by the floret around that situation. The event tree thus obtained is shown in Figure \ref{tree}. This is an example of an asymmetric unfolding of events; if the migrant does not receive an offer to migrate, we do not need to consider whether the offer contains employment. This is denoted here as:
\begin{equation}
    \nexists X_E|  X_O=no.
\end{equation}

\begin{figure}[h]
\centering
    \includegraphics[width=\textwidth]{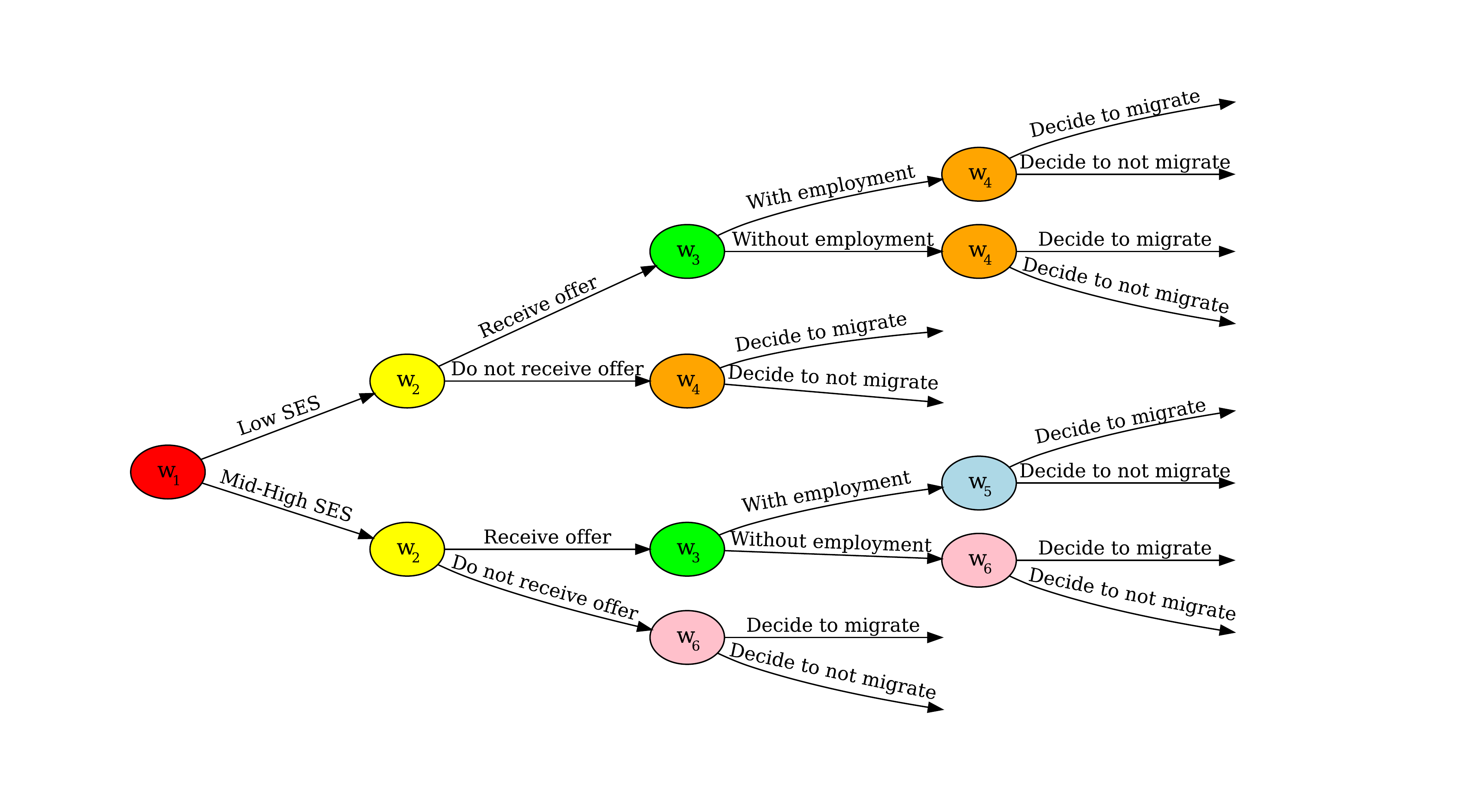}
    \caption{Staged tree representation of the ABM. Here, `SES' refers to socio-economic status. The leaf nodes are suppressed to prevent visual cluttering.}
    \label{staged}
\end{figure}

Next, by looking at the if-then rules within the ABM, we can identify the implicit independence statements that exist within these rules. For the decision rule shown about the decision to migrate, we have the independence statements:
\begin{equation}
    X_M \independent{ } X_O, X_E | X_I=low
\end{equation}

\begin{equation}
    X_M \independent{} X_O | \{X_I=mid\text{-}high , X_E \neq yes\}
\end{equation}
This provides the staging for the CEG. The staging can be represented by a staged tree, an event tree with florets in the same stage coloured the same. The staged tree for this example is shown in Figure \ref{staged}.

For this example, we assume that the other rules in the ABM represent the following statements:
\begin{itemize}
    \item Yellow-- Regardless of socio-economic status, the probability of receiving an offer is the same.
    \item Green-- When an offer is received, the probability of it containing an employment contract is the same, irrespective of socio-economic status.
    \item Orange-- A migrant with low socio-economic status has the same probability of deciding to migrate, irrespective of whether they have received an offer and whether their offer contained an employment contract.
    \item Pink-- A migrant with mid-high socio-economic status has the same probability of deciding to migrate if either (a) they receive an offer but it does not contain an employment contract or (b) they do not receive an offer in the first place.
\end{itemize}
From the staged tree, we can identify the nodes that are in the same position. In this example, $w_4$ and $w_6$ have the same future unfoldings for all future events, and are therefore in the same position.

Note that some nodes are the same stage but not the same position; $w_3$ is one such example, where the probability of the offer having employment is the same but the migrants’ longer-term decision-making will still be influenced by their socio-economic status from earlier in the tree. This example demonstrates a context-specific independence statement: the decision to migrate is independent of whether you have an offer to migrate if your socio-economic status is low.
\begin{figure}[h]
\centering
    \includegraphics[width=\textwidth]{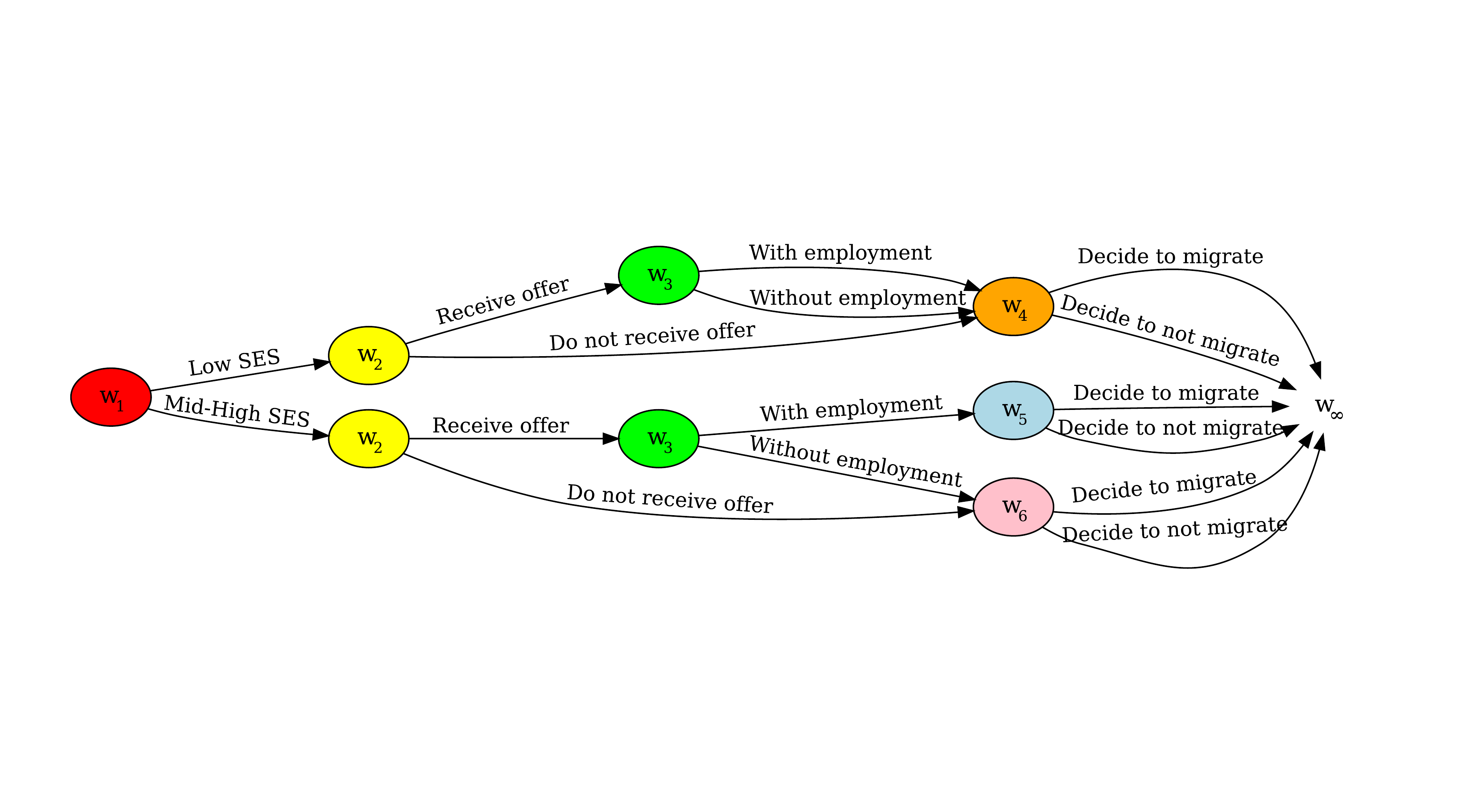}
    \caption{A CEG representation of the above ABM with some examples of independence statements. `SES' stands for socio-economic status.}
    \label{CEG_ABM_example}
\end{figure}

This example shows the CEG can model and provide a compact representation of the conditional independence hypotheses present in the ABM. The transformation from the ABM into the CEG now enables the natural transformation of the model into a Bayesian framework with its associated benefits as described in the previous section. 


\section{Discussion}
\label{Sec:5}

We have demonstrated that we are able to transform ABMs into CEGs. The benefits of this transformation are clear: it provides a compact representation of its independence statements, directly from the topology of the graph. This is valuable in identifying whether the model is making a plausible set of assumptions and making the independence structure accessible to be understood by those without a mathematical background, such as policymakers. The transformation into a CEG also allows for a natural conversion into a Bayesian framework with additional benefits: improved uncertainty quantification, Bayesian inference with available data and Bayesian model selection.

Whilst this paper specifically focuses on migration, CEGs have many potential applications in other domains where ABMs have been used to represent ego-based processes, such as dietary, voting or criminal behaviour.

This research reflects work in process; further investigation is needed to extend this methodology and increase the scope of ABMs that it applies to. Exploration of new representations is ongoing; one extension of the CEG could include the recently developed continuous time dynamic CEG, which is able to accommodate recurrent within the ABM structure and model holding times along the edges between events \cite{shenvi_smith_2020}. Further extensions of interest focus on CEGs which are able to represent the interactions of multi-agent systems players such as in \cite{thwaites_smith_2017}, and agents looping through a CEG with changing probabilities over time depending on previous migration experience. Engaging with this research will provide many avenues of future research to build upon the work presented in this paper, enabling for more full and direct CEG-like representations of an even wider class of ABMs than those discussed above. The full results of this study will be published, alongside any future extensions, in a later paper.



\section*{Acknowledgement}
Peter Strong was supported by the Engineering and Physical Sciences Research
Council and the Medical Research Council grant EP/L015374/1. 
Alys McAlpine was supported by the ESRC grant ES/V006681/1
Jim Q. Smith was funded by the EPSRC [grant number EP/K03 9628/1]. We would like to thank Aditi Shenvi for her valuable comments.
\bibliographystyle{acm}
\bibliography{ref}
\end{document}